\documentclass[%
 aip
 sd,%
 amsmath,amssymb,
 reprint,%
]{revtex4-1}

\draft
\usepackage{graphicx}
\usepackage{color}
\usepackage{dcolumn}
\usepackage{bm}
\usepackage{amsmath}

\newcommand{\rmi}{{\rm i}}
\newcommand{\comm}[1]{}

\begin{document}

\graphicspath{{./figs/}}
%\preprint{}

\title{Multiwall nanotubes of molybdenum disulfide as optical resonators}
\author{D. R. Kazanov$^1$}
\email{kazanovdr@beam.ioffe.ru}
\author{A. V. Poshakinskiy$^1$}
\author{V. Yu. Davydov$^1$}
\author{A. N. Smirnov$^1$}
\author{D. A. Kirilenko$^1$}
\author{M. Rem\v{s}kar$^2$}
\author{S. Fathipour$^3$}
\author{A. Mintairov$^{1,3}$}
\author{A. Seabaugh$^3$}
\author{B. Gil$^{1,4}$}
\author{T. V. Shubina$^1$}

\affiliation{$^1$Ioffe Institute, 26 Politekhnicheskaya, St Petersburg 194021, Russia}
\affiliation{$^2$Jo\v{z}ef Stefan Institute, Jamova cesta 39, 1000 Ljubljana, Slovenia}
\affiliation{$^3$University of Notre Dame, IN 46556, USA}
\affiliation{$^4$L2C, UMR 5221 CNRS-Universit$\acute{e}$ de Montpellier, F-34095, France}

\date{\today}

\begin{abstract}
We study the optical properties of MoS$_2$ nanotubes (NTs) with walls comprising dozens of monolayers. We reveal strong peaks in micro-photoluminescence ($\mu$-PL) spectra when detecting the light polarized along the NT axis. We develop a model describing the optical properties of the nanotubes acting as optical resonators which support the quantization of whispering gallery modes inside the NT wall. The experimental observation of the resonances in $\mu$-PL allows one to use them as a contactless method of the estimation of the wall width. Our findings open a way to use such NTs as polarization-sensitive components of nanophotonic devices.
\end{abstract}

\pacs{Valid PACS appear here}

\maketitle

Optical microcavities of different materials and diverse forms such as planar Bragg resonators with Fabri-P\'erot modes, toroidal resonators or microspheres with whispering gallery modes (WGMs) have been proposed for applications in optoelectronic devices \cite{Vahala2003}. Discovery and studies of the famous carbon nanotubes (NTs) in the last century \cite{Lijima1991,Dresselhaus2000, Rao2002} has opened a new field in material science. Since then, a large variety of inorganic NTs have appeared. They can be divided into two groups depending on the type of bonds between constituent atomic monolayers within a NT wall: covalent/ionic bonds like in conventional three-dimensional materials or van der Waals (vdW) forces. The first group includes materials like ZnO, GaN, SiO$_2$ \cite{Xing2003}. VdW NTs can be made of two-dimensional (2D) monolayers of transition metal dichalcogenides (TMDC) such as MoS$_2$, WS$_2$, WSe$_2$, and some other materials like BN \cite{Tenne1992, Chopra1995, Remskar1996, Remskar2004}. All of them have a form of a hollow cylinder with diameter varying from nanometers to several micrometers and length approaching several millimeters. The NT wall may comprise a single monolayer or several-monolayer wall \cite{Damnjanovic1999}. Wall width of multiwall NTs may approach several dozens of nanometres. Depending on the way how walls roll up into a seamless cylinder, the vdW NTs are classified as armchair, zigzag or chiral. Such folding and inherent intrinsic strain determines the NT electronic properties, e.g., carbon NTs can be either metallic or semiconducting \cite{Wildoer1998}.

The tubular microcavities have attracted much attention due to their interesting polarization properties and applications in nanophotonics, mostly to enhance the light intensity \cite{Wang2014}. Emerging studies of micro- and nanotubes as optical resonators show that they can support WGMs. However, the studies concerned mostly those that were made from thin films or membranes rolled up into tubes. Rolled-up SiO/SiO$_2$ microcavities show resonant modes with the polarization parallel to the tube axis \cite{Huang2009}. The on-chip NTs with InGaN/GaN quantum wells demonstrated lasing controlled by the WGM resonances \cite{Li2017}. 

\begin{figure}[t]
\includegraphics[width=.99\columnwidth]{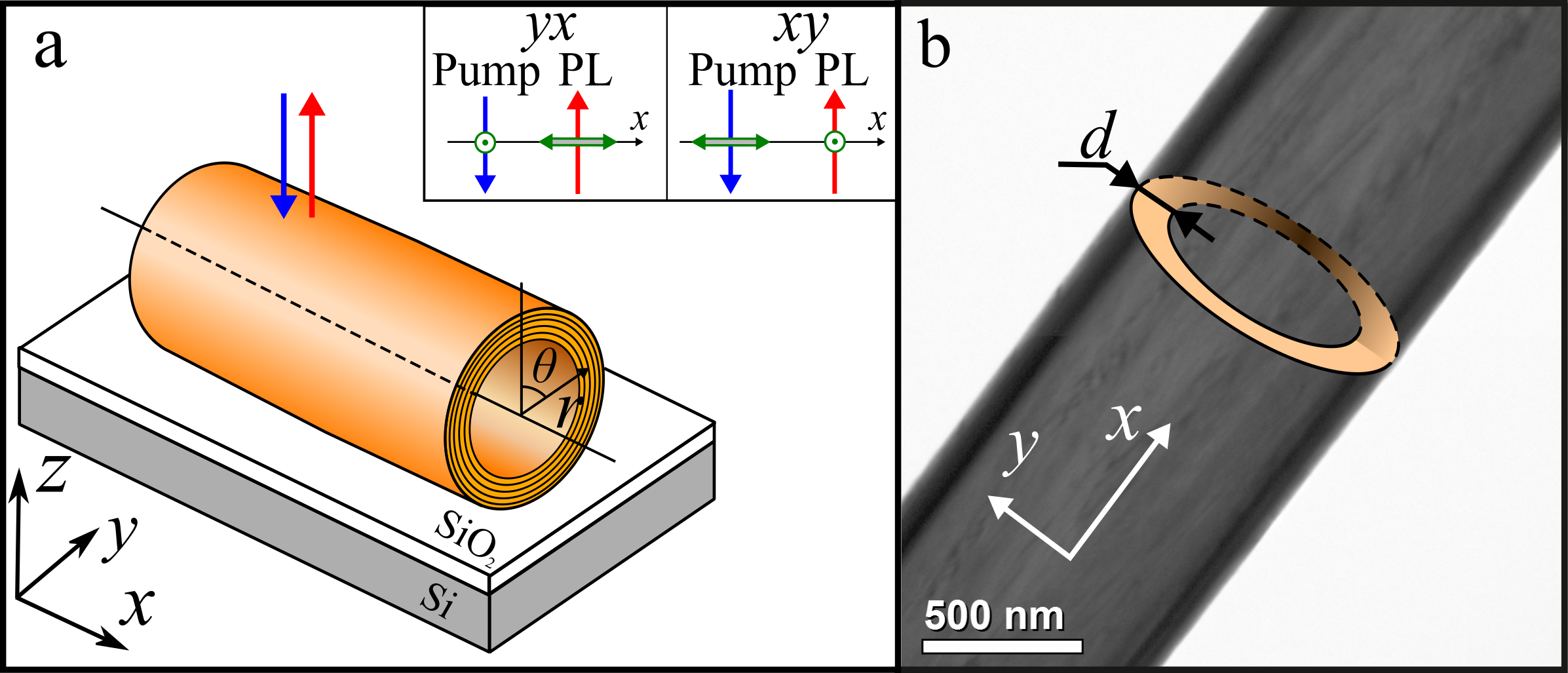}
\caption{ (a) A multiwall MoS$_2$ nanotube on the SiO$_2$/Si substrate. Blue and red lines depict the pump and PL of NT, respectively. Inset shows two different configurations of the $\mu$-PL experiment where green arrows indicates the light polarization. (b) Longitudinal side-view transmission microscopy image of MoS$_2$ NT with the wall width $d \sim$ 50 nm. We depict a cross-section of the NT as a bright ring. The number of monolayers inside the NT wall can be estimated by dividing $d/L$, where $L$ is the interlayer distance in MoS$_2$ stack ($\sim 0.6$ nm).}\label{fig:Sketch} 
\end{figure}

The NTs made of monolayers of TMDC attract attention due to unique properties of parent 2D materials. The TMDC have optical gap and exhibit exciton resonances with strong oscillator strength \cite{Robert2016b}. Their optical properties strongly depend on the number of monolayers \cite{Mak2010}. Defects can affect electronic properties of TMDC as well \cite{Santosh2014}.

The band structure and optical properties of single-wall TMDC NTs have been studied theoretically using calculations based on the density functional theory \cite{Seifert2000a}. It was also predicted that the electronic properties of multiwall NTs of MoS$_2$ should tend to those of the bulk \cite{Seifert2000b}. It was also shown that the band structure, density of states, and vibrational characteristics of NTs depend on strain and folding \cite{Virsek2007,Virsek2009, Zibouche2014, Wang2017}. Currently, the TMDC NTs are becoming promising for future optoelectronic applications. In particular, it was demonstrated that MoS$_2$ multiwall NTs and nanoribbons can be used to create nanodevices such as field-effect transistors with high current densities \cite{Strojnik2014,Fathipour2015}. It was also shown that WS$_2$ NTs could be used as high-perfomance photodetectors \cite{Zhang2012}. To the best of our knowledge, the optical properties of NTs made of 2D vdW 
materials slipped away from the focus of previous research. The luminescence from TMDC nanoscrolls has been only recently reported \cite{Cui2018}.

In this paper, we show the opportunity of using TMDC NTs as optical resonators with strong selection of modes. To elucidate the optical properties of NTs near the frequency of the direct A-exciton transitions we performed micro-photoluminescence ($\mu$-PL) spectroscopy measurements in the range 1.6 - 2.0 eV. They have shown the enhancement of emission by the strong peaks which are polarized along the tube axis $x$. We demonstrate that this selective enhancement is related to the WGM circulating inside the wall of NT. 

\begin{figure}[t]
\includegraphics[width=.99\columnwidth]{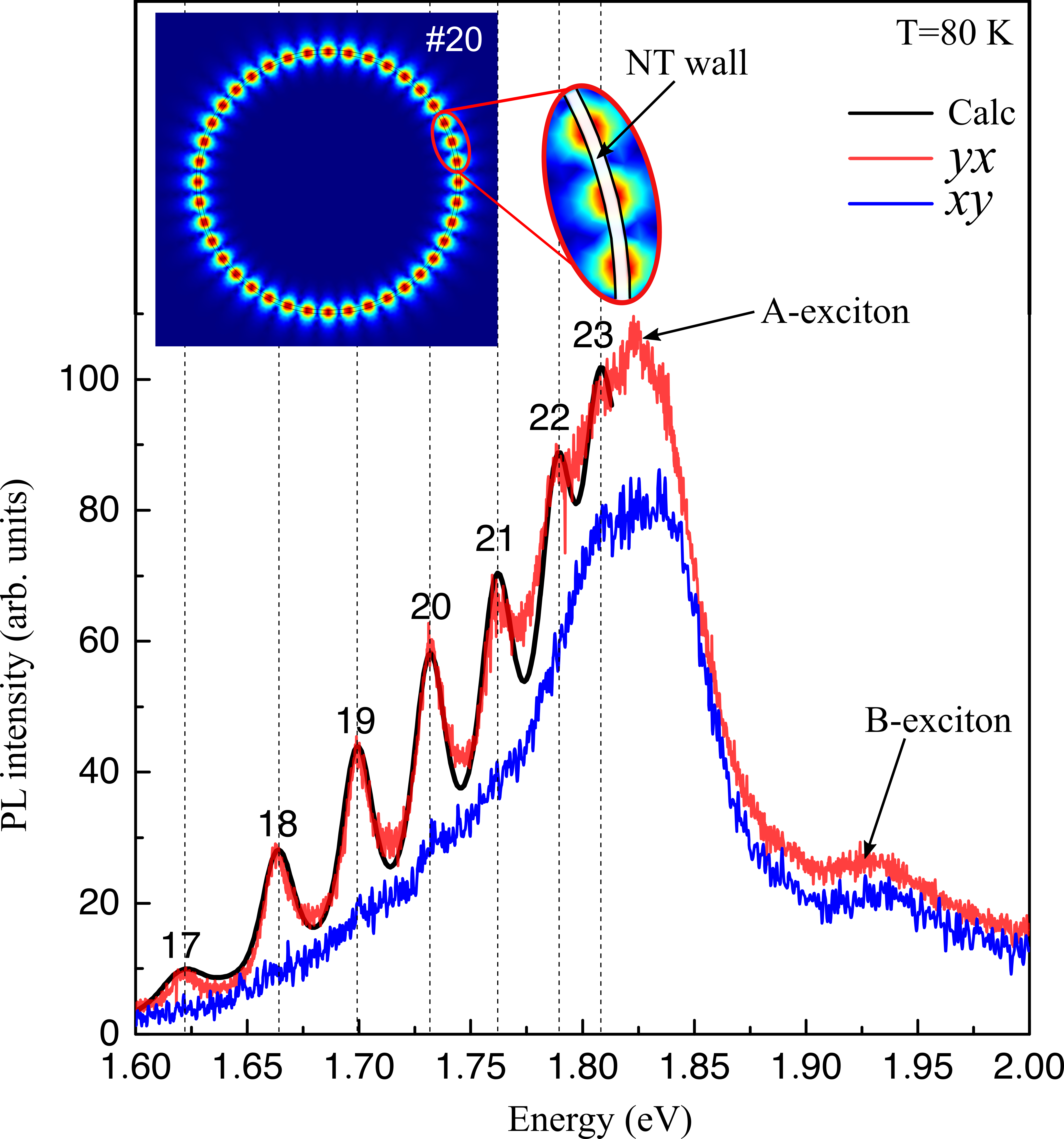}
\caption{Experimental spectra of $\mu$-PL in the $yx$ (red line) and $xy$ (blue line) polarization configurations with 1 mW excitation power. The spectrum in the $yx$ configuration exhibits peaks related to the optical modes. The calculated spectrum for PL in the same configuration is shown by the black line. The angular number of the modes is indicated above the corresponding peak. Inset represents the electric field distribution for $x$-polarized WGM with angular number $m=20$.}\label{fig:PLTM} 
\end{figure}

The studied NTs were grown by chemical transport reaction, using iodine as a transport agent, with very slow rate from the vapor phase that enables to produce NTs with extremely low density of structural defects \cite{Remskar1996,Remskar1998}. For the sake of our demonstration, we have chosen MoS$_2$ NTs which possess brighter emission as compared with WS$_2$ ones, where the lowest excitonic state is dark (see \cite{Malic2018} and references therein), although similar resonant features were observed in these tubes as well. The studied NTs have a form of the hollow cylinders with outer radius $\approx 0.5-1~\mu$m and the wall width $\approx 20-100~$ nm. $\mu$-PL measurements were performed on a Horiba Jobin-Yvon T64000 spectrometer equipped with a confocal microscope, a silicon CCD cooled with liquid nitrogen, and a 600 lines/mm grating. A Nd:YAG-laser line at a wavelength of 532 nm ($\hbar \omega_{{\rm exc}} = 2.33~$eV) was used for $cw$ excitation. The measurements in the temperature range from 80 to 250 K were carried out in a temperature controlled microscope stage Linkam THMS600. A large working distance lens (Mitutoyo 100$\times$NIR (NA = 0.50)) with a spot size of $\approx 2~\mu$m on a sample was used to measure PL from a single NT.

Two different experimental configurations were used in the $\mu$-PL studies (see the inset in Fig.~\ref{fig:Sketch}). In $yx$-configuration, excitation was performed with the light polarization (green arrows) perpendicular to the tube $x$-axis (along $y$) and the light polarized along the $x$-axis (TM-polarized) was detected. In $xy$-configuration, the excitation was $x$-polarized while $y$-polarized PL was detected. 

\begin{figure}[b]
  \includegraphics[width=0.99\columnwidth]{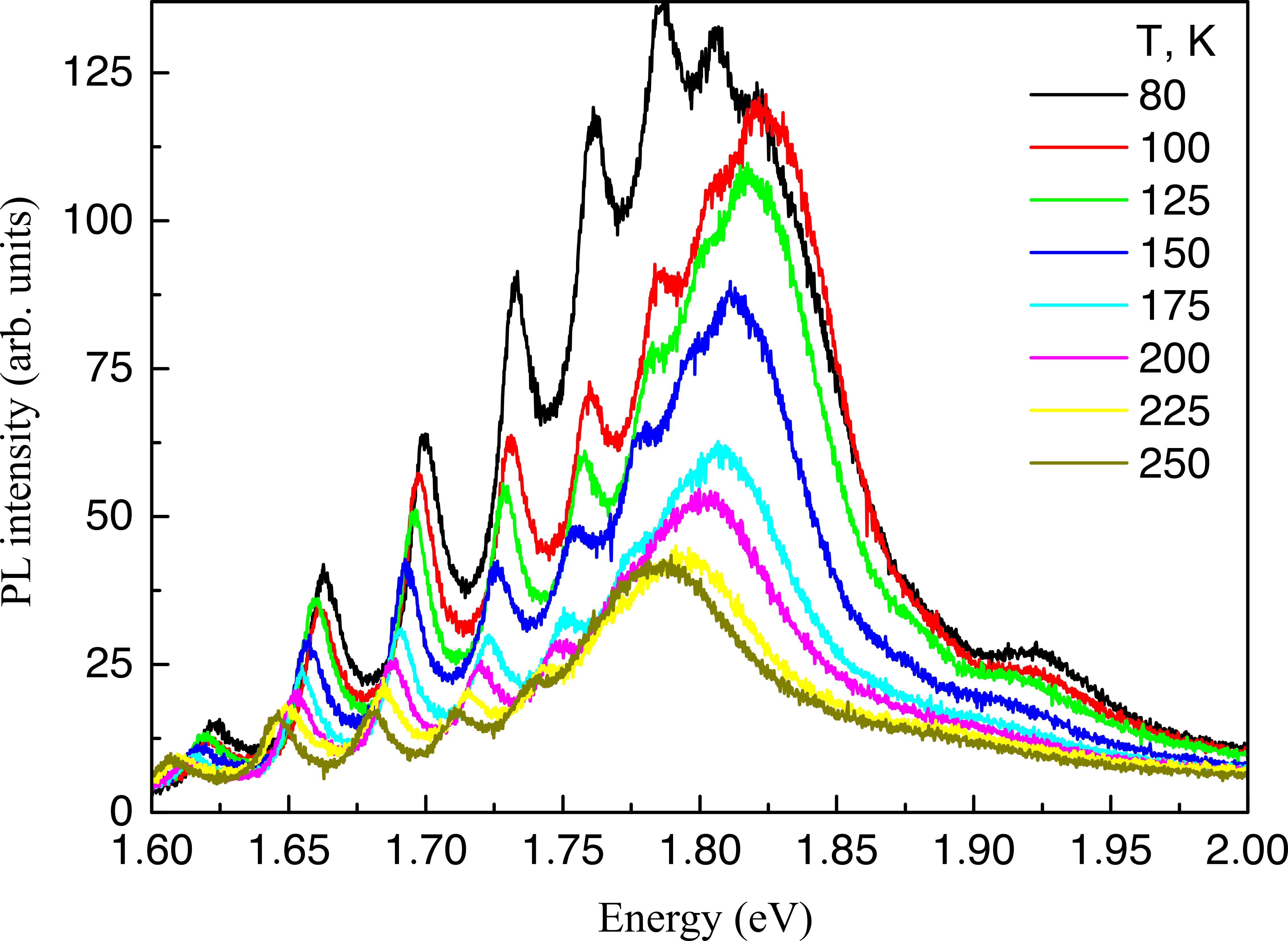}
\caption{Temperature dependent $\mu$-PL spectra in the $yx$-configuration for temperatures varying from 80 K to 250 K.}\label{fig:PLTemp}
\end{figure}

The experimental  $\mu$-PL spectra recorded in these two configurations at 80 K are shown in Fig.~\ref{fig:PLTM}. PL spectra manifest peaks at 1.83 eV and 1.92 eV related to A- and B- excitons, respectively, close to that in planar MoS$_2$ layers \cite{Mak2010}. Both PL spectra exhibit A-exciton emission with a long tail which reveals a sequence of peaks in $yx$-configuration (red line) while in $xy$-configuration (blue line) the peaks are suppressed. The nature of peaks in $yx$-configuration seems to be due to the WGMs inside the NT wall, which contributes to the enhancement of PL intensity which is usually rather weak in the planar multilayer structures due to the indirect nature of the band gap in bulk \cite{Mak2010}. Quality factor, defined as $Q=E/\Delta E$, where $E$ is the energy and $\Delta E$ is the FWHM of resonant peak in spectra, is around 90 for such modes.  Also, the free spectral range (FSR), i.e., the energy splitting between modes, for $x$-polarized modes varies from $\sim$ 50 to 20 meV. The decrease of the FSR with the increase of energy is due to the rise of the refractive index of the material near A-exciton \cite{Zhang2015}. Similar behavior of semiconductor materials occured in other resonators \cite{Shubina2015}.

The temperature dependent measurements in $yx$-configuration (Fig.~\ref{fig:PLTemp}) showed that resonant modes are red-shifted when temperature rises. This is caused by a shrink of the band gap of the material and, thus, the shift of all resonances. Besides, the total PL intensity decreases with temperature due to the increase of the non-radiative exciton decay rate. However, resonant peculiarities are still present in the high-temperature spectra. It means that such NTs can be used as a polarization sensitive-devices or filters even at room temperatures.

To confirm the origin of the PL peaks in the $yx$-configuration and to explain their suppression in the $xy$, we performed the theoretical modeling of the PL spectra of the TMDC NTs. Due to the cylindrical symmetry of the system we describe electromagnetic field in cylindrical coordinates ($x, r,\theta$), where $\theta = 0^{\circ}$ is related to the direction of the incident laser beam. The PL intensity from the NT volume with coordinates $(r,\theta)$ is proportional to the squared amplitude of the incident electric field at that point. Thus, to calculate PL spectra we should find the excitation power inside the NT walls. The electric field inside the wall of the NT can be found by solving a scattering problem of normally incident light on NT, as described in Supplemental Material. The components of the electric field from the incident laser beam that lie in the plane of the NT wall generate excitons with spatial distribution $P^{(y,x)}(r,\theta,\omega_{\rm exc}) \sim |E^{(y,x)}_{x}(r,\theta, \omega _{{\rm exc}})|^2 +  |E^{(y,x)}_{\theta}(r,\theta,\omega_{{\rm exc}})|^2$, where ${\bm E}^{(x)}$ is the electric field inside the NT under $x$-polarized excitation with frequency $\omega_{{\rm exc}}$, ${\bm E}^{(y)}$ is the electric field in the case of $y$-polarized excitation, see Eq.~\eqref{eq:E}. The exciton generation is followed by rapid energy relaxation with the conservation of real space distribution and loss of polarization. The subsequent reemission of photons yields the PL which is modulated by the resonator modes.

\begin{figure}[b]
\includegraphics[width=0.99\columnwidth]{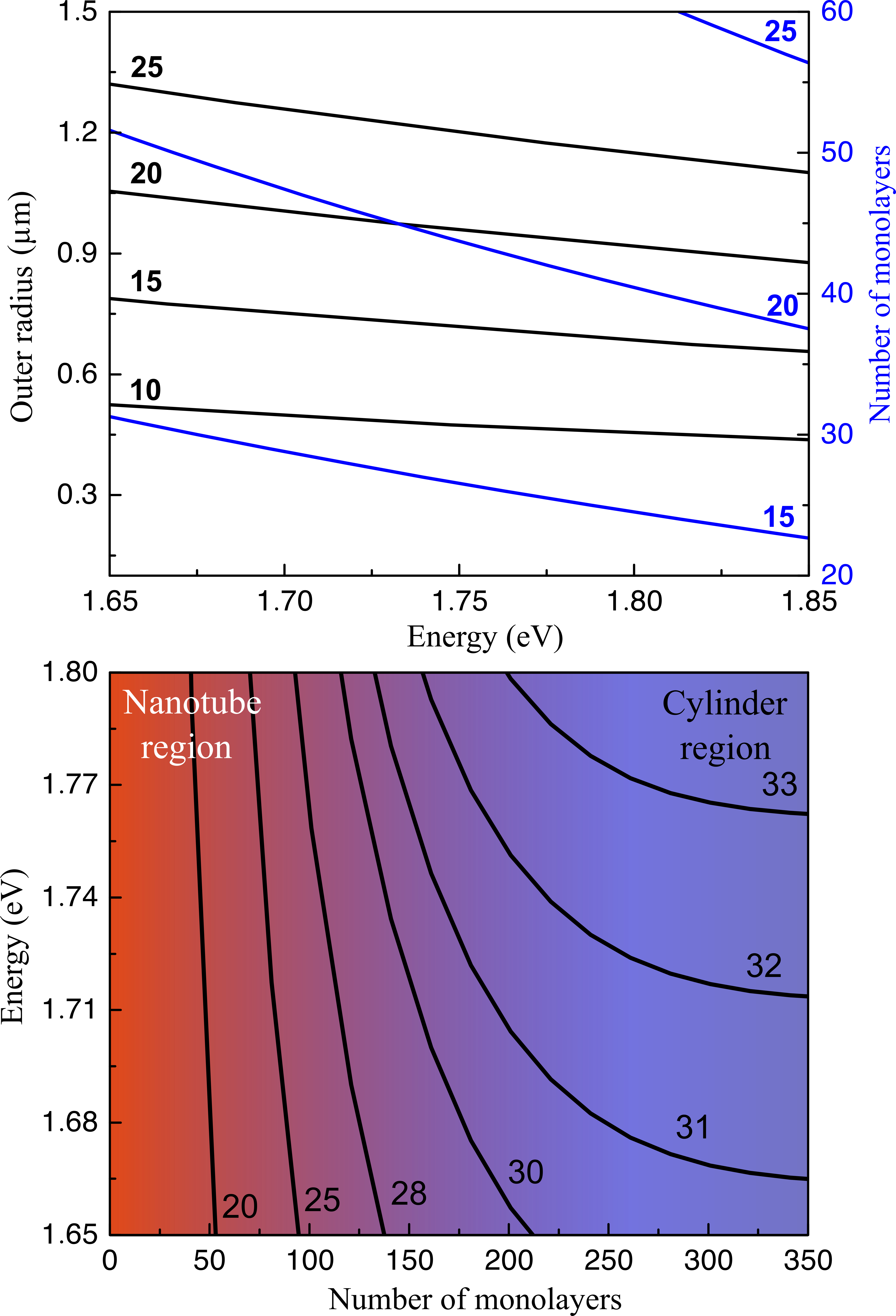}
\caption{(Top) Dependence of the $x$-polarized mode energies on the outer radius of the NT varying from 0.1 to 1.5~$\mu$m for the fixed number of monolayers in the wall $N=45$ (black lines) and on the number of monolayers $N$ varying from 20 to 60 for the fixed outer radius $1~\mu$m (blue lines). (Bottom) Mode energies as a function of number of monolayers varying from 1 to 350 for NT with the fixed outer radius $1~\mu$m. Red
region - light is confined inside the NT wall. Blue region - WGM is localized near outer radius of almost full cylinder.}\label{fig:EvsML}
\end{figure} 
In order to calculate the PL spectra for any polarization of the detected light, we use the Lorentz reciprocity theorem. We replace the problem of light emission from the system in the certain direction by the problem of scattering of the plane wave incident from this direction~\cite{Gippius2005, Iwanaga2007}. The PL spectra for two experimental configurations are then given by
\begin{align}
&{\rm PL}_{yx}(\omega) =\iint |E^{(x)}_x(r,\theta;\omega)|^2 P^{(y)}(r,\theta,\omega_{\rm exc}) dr d\theta ,\\
&{\rm PL}_{xy} (\omega)=\iint |E^{(y)}_{\theta}(r,\theta;\omega)|^2 P^{(x)}(r,\theta,\omega_{\rm exc}) dr d\theta.
\end{align}
Integration is performed over the NT cross-section. These $E^{(x,y)}(r,\theta,\omega)$ fields have the same form as the pump field Eq.~\eqref{eq:E}, but they oscillate at the frequency of detection $\omega$. 

To simulate the real PL spectra one should know not only the electric field distribution that generates excitons in the wall in the real space, but is also mandatory to know the energy distribution of generated excitons. As we see in Fig.~\ref{fig:PLTM}, blue line, the $y$-polarized PL spectrum exhibits negligible resonant peaks; thus, we can use this spectrum as a description of such distribution. We also account for the deviation of the number of monolayers in the NT wall. We relate both the widths and heights of the experimental peaks in terms of a distribution of the wall width of 1 monolayer. The frequency dependent refractive index was taken into account. Its variation was expressed via the combination of the background part and the exciton contribution taking into account the experimental spectra, reported in \cite{Zhang2015}. 

Azimuthal angular numbers ranging from 17 to 23 describe the $x$-polarized WGMs for the tube resonator in the range of energies below A-exciton in MoS$_2$. Modes above 1.8 eV are strongly suppressed due to the absorption induced by the exciton resonance. The electric field distribution of the mode with angular number $m=20$ is shown on the inset of Fig.~\ref{fig:PLTM}. The modeling of $y$-polarized modes has shown that they also can be found in this spectral range, but their angular numbers are lower and vary around 5. In cylindrical cavities the radiative decay and, therefore, WGM $Q$-factor both strongly depend on the azimuthal number. The larger $m$ is, the larger is $Q$-factor \cite{Kaliteevski2007}. This effect is even more pronounced in the tubular geometry, where confinement of the modes in the tube wall is even stronger. Because of that, the $y$-polarized modes with the small angular numbers are not revealed in the PL spectra. For the measured outer radius $1~\mu$m, taking into account the frequency dependent refractive index of MoS$_2$ and the 1-monolayer fluctuation of the wall width, we obtain a perfect agreement between experimental peak positions, widths and amplitudes and calculated ones, assuming the number of monolayers $N=45$ (black line in Fig.~\ref{fig:PLTM}). Thus, such fitting of experimental spectra, which allows to determine the number of monolayers in the wall can be used as a highly accurate contactless method of NTs characterization.

For a deeper insight into the impact of the tube sizes on the optical modes, we calculated the dependencies of the mode energies with the wall width (number of monolayers) and tube radius. Figure~\ref{fig:EvsML} (top) shows the dependence of the $x$-polarized mode energies on the outer NT radius (black lines) for the fixed number of monolayers in the wall $N=45$ and on the number of monolayers (blue lines) for fixed outer radius $1~\mu$m. The modes with angular number $m=10,15,20,25$ that have energies near A-exciton are shown. Alteration of the radius influences the energy of the modes stronger than that of the monolayer number. Further, the smaller the radius is, the smaller is the mode angular number for the same mode energy, thus, the $Q$-factor of such mode is lower and the light weaker interacts with this mode.

Figure~\ref{fig:EvsML} (bottom) demonstrates the transition from the hollow cylinder to the full one. Namely, it shows the dependence of the $x$-polarized mode energies with the number of monolayers in the wall. For the constant outer radius, the increase of the wall width corresponds to the decrease of the inner radius up to the formation of the full cylinder. When the NT wall width is thin (red area), there is a strong dependence of the mode energy on the number of monolayers. This is because the light is confined within the NT wall. The thinner is the wall, the higher is the energy of the mode. With the increase of the monolayer number, the transition to a full cylinder region takes place (blue area). In this case, the WGM is localized in the vicinity of the outer NT radius and the eigenfrequencies become almost independent from the wall width and coincide with those for the full cylinder with the same external radius. The difference between the NT and the cylinder is negligible when the wall width is larger than about $20$ percent of the outer radius.

To summarize, we have shown that the PL spectra of multiwall TMDC NTs comprises strong peaks linearly polarized along the NT axis. We modeled the $\mu$-PL spectra of NTs  for two orthogonal experimental configurations. The perfect agreement with the experimental $\mu$-PL spectra has been obtained taking into account the inhomogeneity of NT parameters and the frequency dependence of the refractive index. We explained the difference between $x$-polarized and $y$-polarized PL spectra by the difference in angular momentum numbers of the corresponding modes in the energy region below A-exciton. We demonstrate that the observed WGMs are confined within the NT wall, between its inner and outer surface. We propose that fitting of the PL spectra modulated by the optical modes can be used as a unique and noninvasive way to estimate the NT wall width. The pronounced PL peaks exist even at room temperature that opens the way to use the TMDC NTs as polarization-sensitive devices.

\begin{acknowledgments}
This work was supported by the Government of the Russian Federation (Project No. 14.W03.31.0011 at the Ioffe Institute).
\end{acknowledgments}

%%%%%%%%%% Merge with supplemental materials %%%%%%%%%%
\pagebreak
\widetext
\begin{center}
\textbf{\large Supplemental material \\ for ``Multiwall nanotubes of molybdenum disulfide as optical resonators"}
\end{center}
%%%%%%%%%% Merge with supplemental materials %%%%%%%%%%
%%%%%%%%%% Prefix a "S" to all equations, figures, tables and reset the counter %%%%%%%%%%
\setcounter{equation}{0}
\setcounter{figure}{0}
\setcounter{table}{0}
\setcounter{page}{1}
\makeatletter
\renewcommand{\theequation}{S\arabic{equation}}
\renewcommand{\thefigure}{S\arabic{figure}}
\renewcommand{\bibnumfmt}[1]{[S#1]}
\renewcommand{\citenumfont}[1]{S#1}
%%%%%%%%%% Prefix a "S" to all equations, figures, tables and reset the counter %%%%%%%%%%

%\subsection*{Maxwell's equations and boundary conditions for NTs}
We solve a scattering problem for the NT with a wall comprising several dozens of monolayers using Maxwell's equations and boundary conditions on the inner radius $r_a$ and the outer radius $r_b$ of the NT. The system has a cylindrical symmetry, thus, one should find the electric and the magnetic field in a form $\tilde {\bm E} ={\bm E} e^{-\rmi (\omega t - k_x x)}$ and $\tilde {\bm H}={\bm H} e^{- \rmi (\omega t - k_x x)}$, where $x$ is along the NT axis. Angular part of the electromagnetic fields should be in a form of $e^{\rmi m \theta}$. We have chosen $E_x$ in three regions of the NT as
\begin{equation}
E_x =
\begin{cases}
A_0 J_m(\sigma r), & 0 \leq r < r_a, \\
A_1 J_m(\kappa r) + A_2 N_m (\kappa r), & r_a \leq r < r_b, \\
A_3 H_m ^ {(2)}(\sigma r) + H_m ^{(1)}(\sigma r), & r_b \leq r,
\end{cases}
\end{equation}
where $\kappa^2=k^2 n_1^2(w) -k_x^2$ and $\sigma^2=k^2 n_0^2 - k_x^2$ , $k=\omega/c$ is the wave vector of light in vacuum. $n_1$ is a frequency-dependent refractive index of the NT wall, $n_0$ is a refractive index of the air. In the first region, the fields should not be infinite at $r=0$, thus, only Bessel function of the first order satisfies the Maxwell's equations. Fields in the second region is described by the superposition of two Bessel functions of the first and second order. Solution in the third region should be taken in a form of the superposition of the incident wave with a single amplitude that is described by a Hankel function of the first order and divergent wave that is described by a Hankel function of the second order. Here, coefficient $A_3$ characterizes a reflection of the incident wave from the NT. We search a solution for magnetic field $H_x$ in the same form as for the electric field.

In cylinder symmetry the Maxwell's equations transform into
\begin{equation}
\begin{aligned}
\frac{d^2 E_x}{d r^2}+\frac{1}{r}\frac{d E_x}{dr}+
\frac{1}{r^2} \frac{d^2 E_x}{d \theta^2}+ [k^2 n(r,\theta)^2 - k_x^2] E_x=0, \\
\frac{d^2 H_x}{d r^2}+\frac{1}{r}\frac{d H_x}{dr}+
\frac{1}{r^2} \frac{d^2 H_x}{d \theta^2}+ [k^2 n(r,\theta)^2 - k_x^2] H_x=0,
\end{aligned} 
\end{equation}
and transverse electromagnetic fields are related to $E_x$ and $H_x$ as follows
\begin{align}
\begin{cases}
  E_r = \frac{\rmi}{[k^2 n^2(r)-k_x^2]}(k_x \frac{dE_x}{dr}+\frac{\omega }{c} \frac{1}{r} \frac{dH_x}{d\theta}), \\
  E_{\theta} = \frac{\rmi}{[k^2 n^2(r)-k_x^2]}(\frac{k_x}{r} \frac{dE_x}{d\theta}-\frac{\omega}{c} \frac{dH_x}{dr}), \\
  H_r = \frac{\rmi}{[k^2 n^2(r)-k_x^2]}(k_x \frac{dH_x}{dr}-\frac{\omega}{c} n^2(r)\frac{1}{r}  \frac{dE_x}{d\theta}),\\
  H_{\theta} = \frac{\rmi}{[k^2 n^2(r)-k_x^2]}(\frac{k_x}{r} \frac{dH_x}{d\theta}+\frac{\omega} {c} n^2(r) \frac{dE_x}{dr}).
\end{cases}
\end{align}
Boundary conditions for the system is continuity of $E_x$, $H_x$, $E_\theta$, $H_\theta$ at $r_a$ and $r_b$. Using all of the above, we obtain matrix for the electromagnetic hybrid waves of the mode $m$: 
%\begin{sideways}
%\begin{minipage}{}
\footnotesize
\[
\begin{pmatrix}\label{matrix:full}
  J_m(\sigma r_a) & -J_m(\kappa r_a) & -N_m(\kappa r_a) & 0 & 0 & 0 & 0 & 0 \\
  0 & -J_m(\kappa r_b) & -N_m(\kappa r_b) & -H_m^{(2)}(\sigma r_b) & 0 & 0 & 0 & 0 \\
   -\frac{\rmi k_x m}{r_a \sigma^2} J_m(\sigma r_a) & \frac{\rmi k_x m}{r_a \kappa^2} J_m(\kappa r_a) & \frac{\rmi k_x m}{r_a \kappa^2} N_m(\kappa r_a)  & 0 & -\frac{\omega}{c \sigma^2} J_m'(\sigma r_a)  & \frac{\omega}{c \kappa^2}  J_m'(\kappa r_a) & \frac{\omega}{c \kappa^2} N_m'(\kappa r_a)  \\
  0 &  -\frac{\rmi k_x m}{r_b \kappa^2} J_m(\kappa r_b) & -\frac{\rmi k_x m}{r_b \kappa^2} N_m(\kappa r_b)  &  -\frac{\rmi k_x m}{r_b \sigma^2} H_m^{(2)}(\sigma r_b) & 0 & -  \frac{\omega}{c \kappa^2} J_m'(\kappa r_b) & - \frac{\omega}{c \kappa^2} N_m'(\kappa r_b)  & -  \frac{\omega}{c \sigma^2} H_m^{'(2)}(\sigma r_b)\\
  0 & 0 & 0 & 0 & J_m(\sigma r_a)& -J_m(\kappa r_a) & -N_m(\kappa r_a) & 0\\
  0 & 0 & 0 & 0 & 0 & J_m(\kappa r_b) & N_m(\kappa r_b) & H_m^{(2)}(\sigma r_b)\\
  \frac{\omega n_0^2}{c \sigma^2} J_m'(\sigma r_a) & - \frac{\omega n_1^2}{c \kappa^2} J_m'(\kappa r_a)  & -  \frac{\omega n_1^2}{c \kappa^2} N_m'(\kappa r_a) & 0 &  \frac{\rmi k_x m}{r_a \sigma^2} J_m(\sigma r_a) & -\frac{\rmi k_x m}{r_a \kappa^2} J_m(\kappa r_a)  & -\frac{\rmi k_x m}{r_a \kappa^2} N_m(\kappa r_a) & 0 \\
 0 & \frac{\omega n_1^2}{c \kappa^2} J_m'(\kappa r_b) &  \frac{\omega n_1^2}{c \kappa^2}  N_m'(\kappa r_b) & -\frac{\omega n_0^2}{c \sigma^2} H_m^{'(2)}(\sigma r_b)  & 0 & \frac{\rmi k_x m}{r_b \kappa^2} J_m(\kappa r_b) & \frac{\rmi k_x m}{r_b \kappa^2} N_m(\kappa r_b)  & -\frac{\rmi k_x m}{r_b \sigma^2} H_m^{(2)}(\sigma r_b) \\
\end{pmatrix}
\]
\normalsize
%\end{minipage}
%\end{sideways}
In the case when $k_x =0$, we obtain a simpler relationship between components of the electromagnetic field with their $x$ projection
\begin{align}
\begin{cases}
  E_r = \frac{\rmi}{k^2 n^2(r)} \frac{\omega }{c} \frac{1}{r} \frac{dH_x}{d\theta}, \\
  E_{\theta} = -\frac{\rmi}{k^2 n^2(r)}\frac{\omega}{c} \frac{dH_x}{dr}, \\
  H_r = -\frac{\rmi}{k^2 n^2(r)} \frac{\omega}{c} n^2(r) \frac{1}{r}  \frac{dE_x}{d\theta}, \\
  H_{\theta} = \frac{\rmi}{k^2 n^2(r)}\frac{\omega} {c} n^2(r) \frac{dE_x}{dr}.
\end{cases}
\end{align}
The matrix splits into two smaller matrixes $S_y$ and $S_x$ that describe pure $y$-polarized mode (TE-mode) and $x$-polarized mode (TM-mode) with angular moment $m$, respectively:
\begin{align}
S = \left( \begin{array}{cc} 
S_{y} & 0 \\
 0 & S_{x}
\end{array}\right), \,
\end{align}
where
\[S_y=
\begin{pmatrix}
J_m(\sigma r_a) & -J_m(\kappa r_a) & -N_m(\kappa r_a) & 0 & \\
  0 & J_m(\kappa r_b) & N_m(\kappa r_b) & - H_m^{(2)}(\sigma r_b)\\
  n_1 J_m'(\sigma r_a) & -n_0 J_m'(\kappa r_a) & - n_0 N_m'(\kappa r_a) & 0\\
  0 & n_0 J_m'(\kappa r_b) & n_0 N_m'(\kappa r_b) & -n_1 H_m^{'(2)}(\sigma r_b)\\
\end{pmatrix}
\]
and
\[S_x=
\begin{pmatrix}
  J_m(\sigma r_a)& -J_m(\kappa r_a) & -N_m(\kappa r_a) & 0\\
  0 & J_m(\kappa r_b) & N_m(\kappa r_b) & -H_m^{(2)}(\sigma r_b)\\
  n_0 J_m'(\sigma r_a)& -n_1 J_m'(\kappa r_a) & -n_1 N_m'(\kappa r_a) & 0\\
  0 & n_1 J_m'(\kappa r_b) & n_1 N_m'(\kappa r_b) & - n_0 H_m^{'(2)}(\sigma r_b)
\end{pmatrix}.
\]
Zero determinant of matrixes $S_y$ and $S_x$ uncovers all resonant frequencies of the system. Using the right part of the linear system which is obtained with respect to a chosen $E_x (H_x)$, we find all frequency dependent coefficients $A$ and, thus, electromagnetic field distribution.

In order to find the electric field from the incident laser beam we use the plane wave expansion in cylindrical coordinates which is a superposition of both incoming and outgoing waves 
\begin{equation}\label{expansion}
e^{i\mathbf{k}\mathbf{r}}=\sum_{m=-\infty}^\infty \frac{\rmi^m}{2} e^{-\rmi m\theta_k} H_m^{(1)}(kr)e^{\rmi m\theta}+\sum_{m=-\infty}^\infty \frac{\rmi^m}{2} e^{-\rmi m\theta_k} H_m^{(2)}(kr)e^{\rmi m\theta}.
\end{equation}

Knowing the coefficients $ A^{(y,x)}$ of the electric field for two polarizations ($y$ or $x$) for mode number $m$ from $S_y$ and $S_x$ matrixes, respectively, and using the plane wave expansion, one can find all components of the electric field inside the wall, excited by the incident laser beam 
\begin{equation}\label{eq:E}
\bm E^{(y,x)}(\omega_{\rm exc}) = \sum_{m} \frac{\rmi^m}{2}[\bm A^{(y,x)}_1 J_m(\kappa r) + \bm A^{(y,x)}_2 N_m(\kappa r)] e^{\rmi m \theta - \rmi \omega_{{\rm exc}} t},
\end{equation}
where the superscript indicates the polarization of light.

\end{document}